\documentclass[a4paper,10pt]{article}

\usepackage{newtxtext}
\usepackage{newtxmath}
\usepackage{graphicx}
\usepackage{xcolor}
\usepackage{booktabs}
\usepackage{multirow}
\usepackage{amsmath}
\usepackage{algorithm}
\usepackage{algpseudocode}
\usepackage{url}
\usepackage{geometry}
\usepackage{titlesec}
\usepackage{caption}
\usepackage{enumitem}
\usepackage{multicol}
\usepackage{fancyhdr}
\usepackage{indentfirst}
\usepackage{flushend}
\usepackage[round,authoryear]{natbib}
\renewcommand{\setemergencystretch}[2]{\emergencystretch=4em}

\fancypagestyle{apiemspages}{%
  \fancyhf{}%
}
\fancypagestyle{apiemsfirstpage}{%
  \fancyhf{}%
  \fancyhead[L]{\fontsize{10}{12}\selectfont\itshape
    Proceedings of the Asia Pacific Industrial Engineering \& Management Systems Conference 2026}%
  \fancyfoot[L]{\fontsize{9}{11}\selectfont
    * These authors contributed equally. \apiemsdag{}: Corresponding Author.}%
}

\titleformat{\section}
  {\normalfont\fontsize{11}{13}\bfseries}{\thesection.}{0.5em}{}
\titleformat{\subsection}
  {\normalfont\fontsize{11}{13}\bfseries}{\thesubsection}{0.5em}{}
\titleformat{\subsubsection}
  {\normalfont\fontsize{11}{13}\bfseries}{\thesubsubsection}{0.5em}{}
\titlespacing*{\section}{0pt}{0.8em}{6pt}
\titlespacing*{\subsection}{0pt}{0.6em}{6pt}
\titlespacing*{\subsubsection}{0pt}{0.45em}{6pt}

\setlist[itemize]{leftmargin=*, itemsep=0pt, topsep=2pt}
\providecommand{\Description}[1]{}

\newcommand{\apiemssup}[1]{\raisebox{0.55ex}{\fontsize{7}{8}\selectfont #1}}
\newcommand{\apiemsdag}{†}
\newcommand{\apiemsblankline}{\par\vspace{8pt}}
\newcommand{\apiemstitleblankline}{\par\vspace{\baselineskip}}
\newenvironment{apiemsabstractblock}
  {\begin{list}{}{\setlength{\leftmargin}{34pt}\setlength{\rightmargin}{34pt}}\item[]}
  {\end{list}}

\begin{document}
\thispagestyle{apiemsfirstpage}

\onecolumn
\begin{center}
{\fontsize{20}{24}\selectfont
HAM-RAG: Hierarchy-Aware Multimodal RAG for\\
Structure-Faithful Interleaved Generation\par}
\apiemstitleblankline
{\fontsize{10}{12}\selectfont
\textbf{Yin Li\apiemssup{*}}\par
The Hong Kong University of Science and Technology (Guangzhou), Guangzhou, China\par
Tel: +86 18689212309, Email: yligt@connect.hkust-gz.edu.cn\par
\apiemsblankline
\textbf{Ziyang Hu\apiemssup{*}}\par
The Hong Kong University of Science and Technology (Guangzhou), Guangzhou, China\par
Tel: +86 18250709675, Email: ziyanghu@hkust-gz.edu.cn\par
\apiemsblankline
\textbf{Zhiyu Guo\apiemssup{*}}\par
The Hong Kong University of Science and Technology (Guangzhou), Guangzhou, China\par
Tel: +86 15557197763, Email: guozhiyu22@mails.ucas.ac.cn\par
\apiemsblankline
\textbf{Xiangyu Liu}\par
The Hong Kong University of Science and Technology (Guangzhou), Guangzhou, China\par
Tel: +86 18866254321, Email: t330034034@mail.bnbu.edu.cn\par
\apiemsblankline
\textbf{Wenbin Li}\par
ASCETEX INTERNATIONAL LIMITED, Hong Kong, China\par
Tel: +86 15021995769, Email: arvinlee@ascetex.com\par
\apiemsblankline
\textbf{Boo-Ho Yang}\par
MOVENSYS Inc., Seongnam-si, Republic of Korea\par
Tel: +86 18129872007, Email: byang@movensys.com\par
\apiemsblankline
\textbf{Rav Lawana}\par
Schneider Electric, Shanghai, China\par
Tel: +86 18616886388, Email: rav.lawana@se.com\par
\apiemsblankline
\textbf{Ziyue Li}\par
Technical University of Munich, Munich, Germany\par
Tel: +49 17679780820, Email: ziyue.li@tum.de\par
\apiemsblankline
\textbf{Wei Zeng}\par
The Hong Kong University of Science and Technology (Guangzhou), Guangzhou, China\par
Tel: +86 18823423978, Email: weizeng@hkust-gz.edu.cn\par
\apiemsblankline
\textbf{Fugee Tsung\apiemssup{\apiemsdag}}\par
The Hong Kong University of Science and Technology (Guangzhou), Guangzhou, China\par
The Hong Kong University of Science and Technology, Hong Kong SAR, China\par
Tel: +86 14714336811, Email: season@hkust-gz.edu.cn\par
}
\end{center}

\begin{apiemsabstractblock}
\noindent\textbf{Abstract.}
Existing multimodal RAG methods often flatten structured documents into isolated text and image units, weakening the source organization and local text-image logic needed for faithful evidence selection and placement. We propose HAM-RAG, a Hierarchy-Aware Multimodal RAG framework for structure-faithful interleaved generation. HAM-RAG uses document hierarchy as a grounding signal across retrieval and generation, contextualizing textual and visual evidence and preserving source position and local text-image relations in the prompt. We further introduce HAM-Bench, covering Wukong, Wiki, arXiv, and Recipe across game walkthroughs, web pages, scientific papers, and step-wise recipe documents. Across multiple backbones, HAM-RAG improves the main multimodal average by 17.3\% over the strongest non-hierarchical baseline. On Wukong, HAM-RAG improves Img-CBS by 24.2\% over the strongest non-hierarchical baseline, demonstrating substantially better local text-image alignment. The main experiments and ablation study together demonstrate that document hierarchy is a key grounding signal for faithful image selection, placement, and local text-image alignment. These findings highlight the value of hierarchy-aware grounding for reliable multimodal assistants that generate answers faithful to the source organization, procedural structure, and local text-image evidence of structured documents, such as technical manuals, maintenance guides, and industrial SOPs. The code is available at \url{https://github.com/MCCodeAI/HAM-RAG.git}.

\vspace{\baselineskip}
\end{apiemsabstractblock}
\vspace{0.8em}

\begin{multicols}{2}

\section{INTRODUCTION}

Retrieval-Augmented Generation (RAG) enhances large language models by grounding generation in external knowledge sources~\citep{lewis2020retrieval}. Early multimodal RAG systems retrieve textual and visual evidence to support answer generation~\citep{chen2022murag}. More recent approaches and benchmarks further study multimodal and interleaved text-image responses~\citep{zhu2025murar,ma2024m2rag,yu2025mramgbench,dong2025mmdocrag}. Yet many real-world multimodal sources are \emph{structured documents}, not flat pools of independent passages and images. Scientific papers organize figures under sections and captions; recipes arrange images around procedural steps; game walkthroughs and web pages rely on headings, local descriptions, and ordered visual demonstrations. In these documents, hierarchy is not cosmetic metadata. It defines the semantic scope of a text span, the local context that makes an image relevant, and the order in which multimodal evidence should be understood.

The goal of multimodal RAG in this setting is therefore not only to retrieve useful text and images, but to preserve the source document's contextual organization and text-image logic in the generated answer. This is particularly important for interleaved generation. An image can be topically related to the query but still be wrong if it is taken from a different section, a different step, or a different local explanation. Conversely, a correct image can become misleading when inserted beside unsupported text. As shown in Figure~\ref{fig:qualitative_examples}, text-only generation lacks visual grounding, while a hierarchy-agnostic multimodal baseline can select plausible visual evidence but arrange it with weak local alignment. A structure-faithful multimodal RAG system should select images and place them where their surrounding text is supported by the original document.

This paper studies whether document hierarchy should be treated as a first-class grounding signal for multimodal RAG over structured documents. We focus on three coupled requirements: \emph{textual scope}, where retrieved text should retain the section or step context that gives it meaning; \emph{image grounding}, where selected images should be tied to their captions, headings, and neighboring descriptions; and \emph{image placement}, where generated answers should preserve the local relation between visual evidence and the text it supports. This formulation differs from generic multimodal retrieval because the correctness of an answer depends on source-local organization, not just global semantic relevance.\par

\columnbreak
\begin{center}
    \includegraphics[width=0.90\columnwidth]{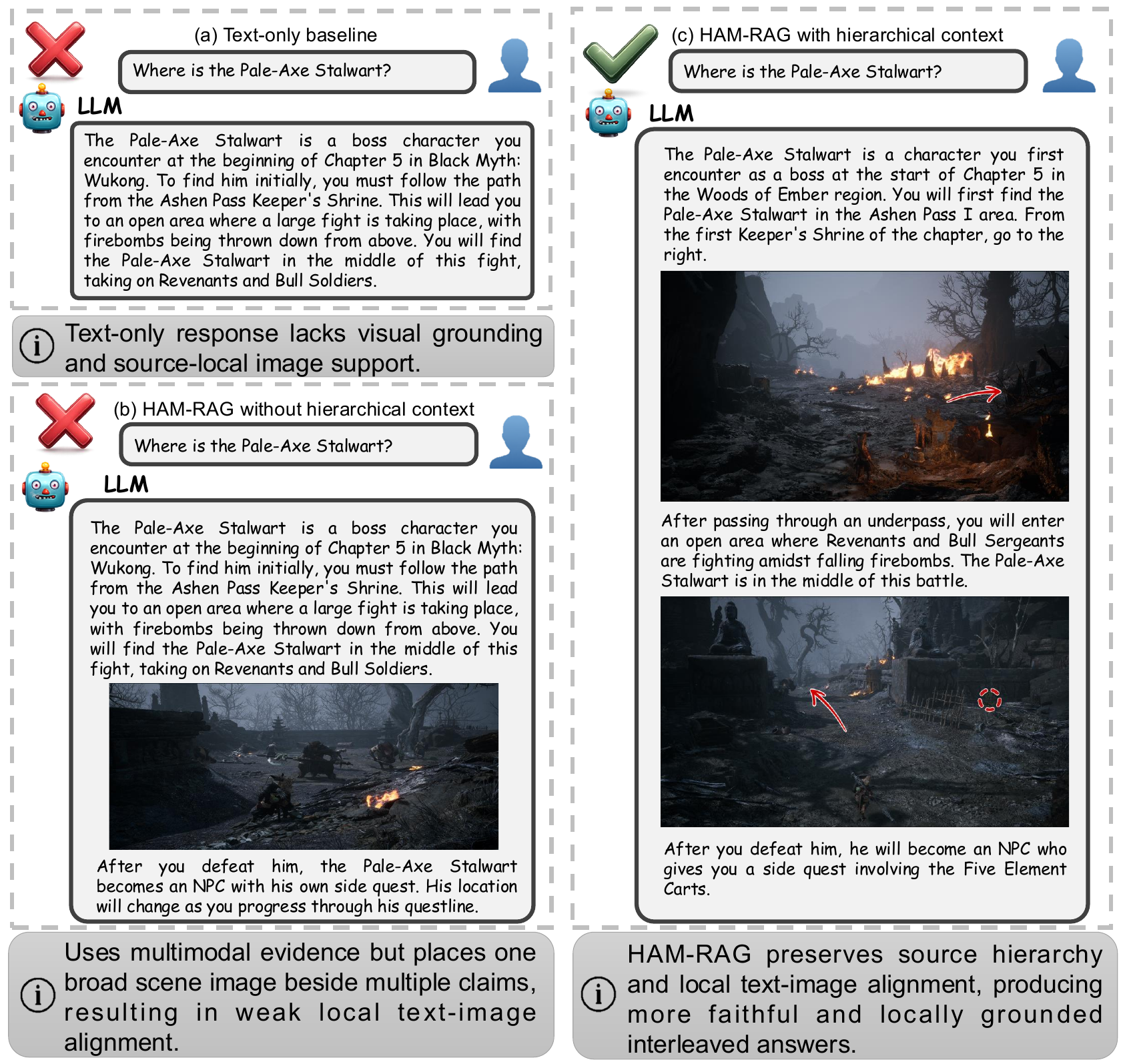}
    \captionsetup{justification=justified,singlelinecheck=false}
    \captionof{figure}{Comparison of a text-only baseline, a flat RAG baseline, and HAM-RAG. HAM-RAG better preserves source hierarchy and local text-image alignment in its final answers.}
    \Description{Qualitative comparison showing a structured document, a question, and generated answers from text-only, hierarchy-agnostic multimodal, and HAM-RAG systems.}
    \label{fig:qualitative_examples}
\end{center}

To address this problem, we propose HAM-RAG, a Hierarchy-Aware Multimodal Retrieval-Augmented Generation framework for structured documents. HAM-RAG represents textual and visual units with hierarchy-aware contextual signals. Specifically, on the text side, each retrievable unit is enriched with source-grounded heading and section context. On the image side, each image is represented with its document position, heading context, nearby text, caption or alt text, and a context-aware visual description. These representations allow retrieval to operate over evidence units that expose their structural role and local cross-modal context. The generator then receives a structured prompt containing ranked text and image evidence, source identifiers, document positions, and image descriptions, encouraging the final answer to use images in locally supported contexts.

We further introduce HAM-Bench, a benchmark for evaluating hierarchy-aware multimodal RAG. HAM-Bench covers diverse structured document types, including web pages, scientific papers, step-wise recipe documents, and game walkthroughs. It includes both natively structured documents and documents whose structure is reconstructed from available source materials. Unlike existing multimodal QA and RAG benchmarks~\citep{chang2022webqa,li2022mmcoqa,yu2025mramgbench,dong2025mmdocrag}, HAM-Bench focuses on structure-preserving multimodal generation and provides text-image local alignment metadata for evaluating whether generated visual evidence is selected and placed in the correct document context. Based on HAM-Bench, we evaluate hierarchy-aware multimodal RAG from multiple perspectives, including retrieval sufficiency, image selection, image-context alignment, and overall answer quality.

Experiments show that HAM-RAG improves the main multimodal average by 17.3\% across matched backbones and increases Img-CBS by 24.2\% on Wukong over the strongest non-hierarchical baseline. These gains demonstrate that hierarchy-aware modeling strengthens visual grounding and local text-image alignment while preserving source organization.

Our contributions are:
\begin{itemize}[leftmargin=1.1em,labelsep=0.35em,itemsep=0.25\baselineskip,topsep=2pt,parsep=0pt,partopsep=0pt]
    \item We formulate structure-faithful multimodal RAG and propose HAM-RAG, which uses document hierarchy to contextualize text and image evidence.
    
    \item We construct HAM-Bench, covering native and reconstructed structured documents with local text-image alignment metadata.
    
    \item We design an evaluation protocol and show consistent gains in grounding and structural faithfulness across datasets and backbones.
\end{itemize}

\section{RELATED WORK}
\label{sec:related_work}

Multimodal retrieval-augmented generation extends text-only RAG to settings in which textual and visual evidence are retrieved to support answer generation. Early systems such as MuRAG~\citep{chen2022murag} retrieve both textual and visual knowledge to support text answer generation. More recent approaches and task formulations such as MuRAR~\citep{zhu2025murar} and M2RAG~\citep{ma2024m2rag} extend this setting toward multimodal answers, while VisRAG~\citep{yu2024visrag} preserves page appearance and layout through visual document retrieval. However, many existing approaches emphasize retrieval architecture, evidence fusion, or generation without explicitly modeling document hierarchy and local text-image relations. They therefore provide limited support for structured documents, where the meaning of an evidence unit often depends on its hierarchical position and surrounding context.

Structure-aware retrieval and multimodal document understanding address some limitations of flat chunking. Wiki-LLaVA uses hierarchical retrieval over multimodal Wikipedia documents~\citep{caffagni2024wikillava}; M3DocVQA introduces the M3DocRAG framework, which preserves page-level visual information during cross-page and multi-document retrieval~\citep{cho2024m3docrag}; and MLDocRAG organizes multimodal cross-page evidence through a query-centric graph~\citep{zhang2026mldocrag}. These studies show that structural and cross-page context is important for document-level understanding. Nevertheless, their primary focus is document QA, long-document retrieval, or multimodal understanding rather than preserving source organization and local text-image logic during interleaved generation. HAM-RAG instead treats document hierarchy not merely as auxiliary metadata, but as a core grounding signal for both textual and visual evidence.

Benchmarks for multimodal RAG and document QA evaluate how systems use multimodal evidence and generate grounded answers. WebQA~\citep{chang2022webqa} and MMCoQA~\citep{li2022mmcoqa} evaluate question answering grounded in multimodal evidence, whereas MRAMG-Bench~\citep{yu2025mramgbench} and MMDocRAG~\citep{dong2025mmdocrag} extend evaluation to multimodal retrieval, evidence selection, and interleaved text-image answer generation. These benchmarks do not explicitly measure whether outputs preserve section hierarchy, procedural order, and local correspondence between text and images. HAM-Bench fills this gap by evaluating hierarchy-aware multimodal RAG over structure-preserving documents, with metrics for grounding, image placement, local text-image alignment, and structural faithfulness to the source.

\section{HAM-RAG FRAMEWORK}
\label{sec:method}

HAM-RAG performs retrieval-augmented generation over hierarchical multimodal documents. As illustrated in \mbox{Figure~\ref{fig:figure-method}}, it consists of an offline hierarchy-aware indexing stage and an online multimodal retrieval-and-generation stage. Unlike flat RAG pipelines that represent passages and images as isolated chunks, HAM-RAG uses document hierarchy as a shared signal for evidence representation, retrieval, and prompt construction.

\subsection{Hierarchy-Aware Document Representation}

Let \(\mathcal{D}=\{D_1,\ldots,D_N\}\) denote a corpus of structured multimodal documents. Each document is parsed into a hierarchy tree
\begin{equation}
T_D=(V_D,E_D),
\end{equation}
where internal nodes represent the document, sections, subsections, or procedural steps, leaf nodes represent retrievable text and image units, and edges encode parent--child relations. The hierarchy is recovered from available document markup, heading levels, numbering patterns, figure--caption links, layout cues, and source order.

For each retrievable leaf unit, HAM-RAG constructs a structured evidence object
\begin{equation}
e_i=\{x_i,m_i,p_i,a_i,l_i,r_i\},
\end{equation}
where \(x_i\) is the original text or image content, \(m_i\) is its modality, \(p_i\) is the path from the document root to the unit, \(a_i\) contains ancestor headings and section context, \(l_i\) contains nearby text-image context, and \(r_i\) records the unit's local relation to surrounding evidence.

\end{multicols}
\twocolumn[{%
\begin{center}
    \includegraphics[width=0.95\textwidth]{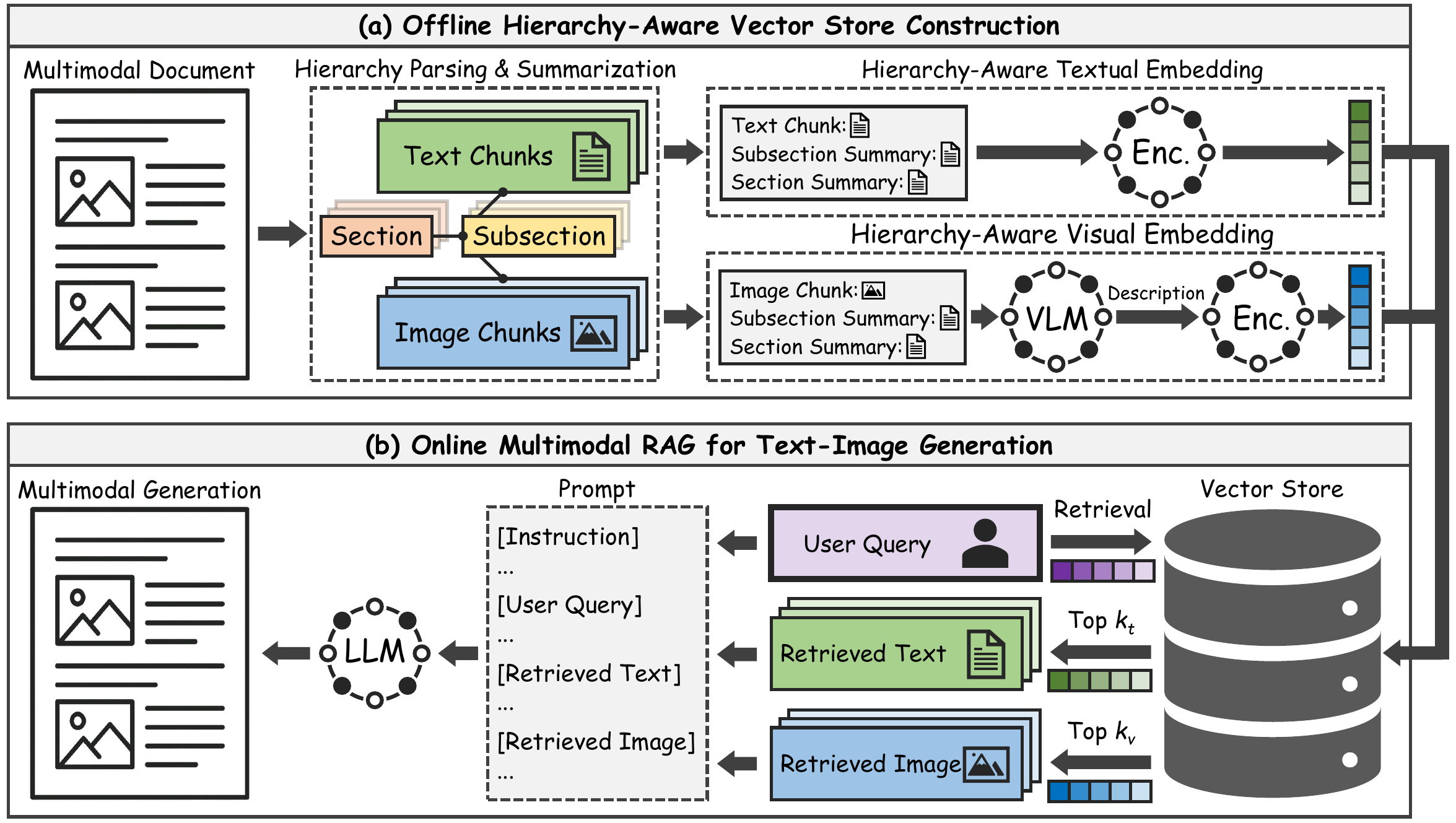}
    \captionsetup{justification=justified,singlelinecheck=false}
    \captionof{figure}{Overview of HAM-RAG. The offline stage constructs and indexes hierarchy-aware text and image evidence, while the online stage retrieves ranked evidence and organizes it for structure-faithful interleaved generation.}
    \Description{Pipeline diagram of HAM-RAG, including offline hierarchy parsing, hierarchy-aware text and image evidence construction, vector indexing, online retrieval, prompt assembly, and interleaved generation.}
    \label{fig:figure-method}
\end{center}
}]

\noindent For example, \(r_i\) may indicate that an image illustrates the current procedural step, belongs to a figure caption, or supports the preceding paragraph. This representation makes the semantic scope and local role of each evidence unit explicit before retrieval.

\subsection{Offline Evidence Construction and Indexing}

\mbox{Figure~\ref{fig:figure-method}(a)} summarizes the offline stage. For a text unit, HAM-RAG serializes the original content together with its document title, heading path, section context, and neighboring evidence into a hierarchy-aware textual representation. This preserves source wording while exposing the structural context needed to distinguish passages that are topically similar but belong to different sections or steps.

For an image unit, the framework retains its source position, heading path, caption or alt text, and nearby textual context. It additionally generates a context-aware visual description conditioned on both the image and its surrounding document evidence. This description supplements sparse or missing captions and helps connect visual content to the section or procedure in which it appears. Text and image evidence are then serialized into field-structured embedding prompts, embedded, and stored in a shared vector store together with provenance metadata such as document identifier, source position, modality, and image reference. Consequently, hierarchy affects retrieval through the indexed representation rather than being appended only after retrieval.

\subsection{Online Retrieval and Interleaved Generation}

As shown in \mbox{Figure~\ref{fig:figure-method}(b)}, the online stage receives a user query and retrieves the top-ranked textual and visual evidence under separate budgets \(k_t\) and \(k_v\). HAM-RAG retains each hit's relevance score, source identifier, document position, hierarchy-aware content, and local text-image relation. Keeping these fields allows evidence from different documents or sections to remain distinguishable during generation.

The selected evidence is assembled into a structured prompt
\begin{equation}
P_{\mathrm{ham}}=\operatorname{Assemble}(q,\mathcal{E}^{*}),
\end{equation}
where \(\mathcal{E}^{*}\) contains the ranked text and image evidence. Text blocks expose source-grounded headings and neighboring context, while image blocks include the image reference, caption or alt text, source position, and context-aware description. The generator therefore receives not only relevant content but also the local organization needed to decide which image supports which statement and where that image should be placed. In this way, hierarchy participates twice: it shapes the retrieval representation offline and guides evidence organization and interleaved text-image generation online.

\section{HAM-BENCH}
\label{sec:ham_bench}

\begin{table}[t]
    \centering
    \scriptsize
    \setlength{\tabcolsep}{1.8pt}
    \caption{Comparison of HAM-Bench with representative multimodal QA/RAG benchmarks.}
    \label{tab:benchmark_comparison}
    \resizebox{\columnwidth}{!}{%
    \begin{tabular}{l l c c c c c}
\toprule
\textbf{Benchmark}
& \textbf{Setting}
& \textbf{MM}
& \textbf{Inter.}
& \textbf{Struct.}
& \textbf{Recon.}
& \textbf{Hier. Eval.} \\
\midrule
WebQA~\citep{chang2022webqa}
& MM QA
& Yes
& No
& No
& No
& No \\
MRAMG-Bench~\citep{yu2025mramgbench}
& MM RAG
& Yes
& Yes
& Partial
& Partial
& No \\
MMDocRAG~\citep{dong2025mmdocrag}
& MM DocRAG
& Yes
& Yes
& Partial
& No
& No \\
HAM-Bench
& Hier. MM RAG
& Yes
& Yes
& Yes
& Yes
& Yes \\
\bottomrule
    \end{tabular}%
    }
\end{table}

\begin{table*}[t]
    \centering
    \small
    \setlength{\tabcolsep}{3.8pt}
    \caption{Structural statistics of HAM-Bench. Native/Recon. indicate native/reconstructed document hierarchies. Imgs counts unique source images; Img/Ans, Img/Doc, and QA/Doc are mean images per answer, image occurrences per document, and QA-document associations per document, respectively. \#H1/\#H2 are mean level-1/2 headings per document.}
    \label{tab:ham_bench_stats}
    \resizebox{\textwidth}{!}{%
    \begin{tabular}{l l c c c c c c c c c c c}
\toprule
\textbf{Subset} & \textbf{Domain} & \textbf{Struct. Source} & \textbf{Multi-Doc} & \textbf{Proc. QA} & \textbf{Docs} & \textbf{QAs} & \textbf{Imgs} & \textbf{Img/Ans} & \textbf{Img/Doc} & \textbf{QA/Doc} & \textbf{\#H1} & \textbf{\#H2} \\
\midrule
Wukong & Game Walkthrough & Native & Yes & Yes & 431 & 540 & 2594 & 4.60 & 8.83 & 2.11 & 1.04 & 2.03 \\
Wiki & Web & Recon. & No & No & 538 & 500 & 539 & 1.00 & 0.98 & 0.93 & 1.00 & 3.65 \\
arXiv & Academic & Recon. & No & No & 101 & 191 & 3136 & 1.35 & 31.05 & 1.89 & 1.02 & 12.78 \\
Recipe & Lifestyle & Recon. & No & Yes & 1528 & 2360 & 8569 & 2.15 & 5.60 & 1.54 & 1.00 & 4.97 \\
\bottomrule
    \end{tabular}%
    }
\end{table*}

In Table~\ref{tab:benchmark_comparison}, MM denotes multimodal evidence, Inter. denotes interleaved text-image answer generation, Struct. denotes explicit preservation or use of document structure, Recon. denotes reconstruction of hierarchy from flattened sources, and Hier. Eval. denotes explicit hierarchy-aware evaluation. We assign Yes when a benchmark fully supports a feature, Partial when the support is limited or indirect, and No when the feature is absent.

To evaluate whether multimodal RAG can faithfully preserve the contextual organization and text-image logic of structured documents, we introduce HAM-Bench, a benchmark built on hierarchy-preserving multimodal corpora. 
Unlike prior multimodal QA/RAG benchmarks that mainly focus on evidence usage or multimodal answer quality, HAM-Bench is designed to test whether retrieval and generation can exploit document hierarchy, local context, and text-image associations under a shared retrieval-and-generation setting.

As shown in Table~\ref{tab:benchmark_comparison}, HAM-Bench differs from existing benchmarks in three aspects. 
First, it uses multimodal documents whose hierarchical organization is explicitly preserved. 
Second, when source documents are originally flattened in prior resources, we reconstruct their document boundaries, section structures, and local text-image alignments. 
Third, it supports evaluation of hierarchy-aware retrieval and interleaved text-image generation, making it suitable for studying structure-aware multimodal RAG.

\subsection{Benchmark Construction and Statistics}

{\tolerance=3000
\emergencystretch=1em
\hyphenpenalty=10000
\exhyphenpenalty=50
HAM-Bench contains four diverse types of structured documents. 
\textit{Wukong} is built from publicly available \textit{Black~Myth:~Wukong} walkthroughs, which naturally contain section-level hierarchy and dense interleaved visual evidence. 
\textit{Wiki} is traced back to WikiWeb2M~\citep{burns2023wikiweb2m} pages, \textit{arXiv} to original PDFs parsed into section-structured \mbox{Markdown} using MinerU~\citep{wang2024mineru}, and \textit{Recipe} to source recipe documents reorganized into step-wise hierarchical procedures. 
For all reconstructed subsets, we restore structures already present in the original sources, including headings, section boundaries, figure-caption links, and procedural steps, rather than redesigning the content.
\par}

Table~\ref{tab:ham_bench_stats} shows that the four subsets pose complementary structural challenges. \textit{arXiv} has deep section hierarchies and dense visual content; \textit{Recipe} and \textit{Wukong} emphasize procedural order and local text-image alignment, with \textit{Wukong} additionally requiring multi-document reasoning; and \textit{Wiki} contains shallower hierarchies and fewer images. Together, they cover scientific, procedural, and web documents.

\section{EXPERIMENTS}
\label{sec:experiments}
\vspace{-0.3\baselineskip}

\subsection{Experimental Setup}
\label{ssec:setup}
\vspace{-0.3\baselineskip}

\subsubsection{Compared Methods}
\vspace{-0.3\baselineskip}

We adopt and reimplement the three baseline paradigms used in MRAMG-Bench---LLM-Based, MLLM-Based, and Rule-Based---because MRAMG-Bench shares our goal of retrieval-augmented interleaved text-image generation~\citep{yu2025mramgbench}. In contrast, related systems such as VisRAG, Wiki-LLaVA, and M3DocRAG primarily target retrieval or document QA rather than interleaved text-image answer generation, making them not directly comparable under our generation metrics. All methods are evaluated under the same retrieval-and-generation setting to isolate the effect of document hierarchy.

\begin{itemize}[leftmargin=1.1em,labelsep=0.35em,itemsep=0pt,topsep=0pt,parsep=0pt,partopsep=0pt]
    \item HAM-RAG.
    Our proposed hierarchy-aware framework. 
    It enriches both text and image units with structural context derived from document hierarchy before retrieval, so that retrieved evidence better preserves the source document's contextual organization and local text-image associations.

    \item LLM-Based.
    A text-only generation paradigm that represents images through captions and surrounding textual descriptions. 
    The generator receives retrieved text together with image textual proxies, but does not directly process visual inputs.

    \item MLLM-Based.
    A multimodal generation paradigm that directly feeds retrieved texts and selected images into an MLLM. 
    Since MLLMs are constrained by the number of input images, a CLIP-based filtering module is used to retain the most query-relevant images before generation.

    \item Rule-Based.
    A post-hoc image insertion paradigm. 
    It first generates a pure textual answer and then inserts images by matching candidate images with answer sentences using lexical and semantic similarity.
\end{itemize}

For each paradigm, we instantiate the generation module with multiple backbone models when applicable, including GPT-5, Gemini-2.5, and Qwen2.5, among others. 
This allows us to examine whether the effect of hierarchy-aware modeling remains consistent across different model families.

\begin{table*}[t]
    \centering
    \caption{Main multimodal generation results on HAM-Bench, covering image selection, local image-context alignment, and holistic multimodal quality. MM Avg. is the unweighted mean of Img-F1, Img-CBS, and Qual. $\times 20$ across all subsets.}
    \label{tab:overall_performance}
    \resizebox{\textwidth}{!}{%
    \begin{tabular}{@{}cc|ccc|ccc|ccc|ccc|c@{}}
\toprule
\multirow{2}{*}[-2pt]{\textbf{Framework}} 
& \multirow{2}{*}[-2pt]{\textbf{Model}} 
& \multicolumn{3}{c|}{\textbf{Wukong $\uparrow$}} 
& \multicolumn{3}{c|}{\textbf{Wiki $\uparrow$}} 
& \multicolumn{3}{c|}{\textbf{arXiv $\uparrow$}} 
& \multicolumn{3}{c|}{\textbf{Recipe $\uparrow$}} 
& \textbf{MM Avg. $\uparrow$} \\
\cmidrule(lr){3-5} \cmidrule(lr){6-8} \cmidrule(lr){9-11} \cmidrule(lr){12-14}
& 
& \textbf{Img-F1} & \textbf{Img-CBS} & \textbf{Qual.}
& \textbf{Img-F1} & \textbf{Img-CBS} & \textbf{Qual.}
& \textbf{Img-F1} & \textbf{Img-CBS} & \textbf{Qual.}
& \textbf{Img-F1} & \textbf{Img-CBS} & \textbf{Qual.}
& \\
\midrule
\multirow{9}{*}{HAM-RAG}
& DeepSeek-V3             &55.67&55.49&3.74&\textbf{100}&70.48&3.39&50.57&50.82&3.34&74.81&65.74&3.76 &67.35\\
& Gemini-2.5-Flash        &65.31&59.17&3.76&97.53&69.71&3.39&54.01&49.24&3.39&87.82&71.10&3.81 &70.07\\
& Gemini-2.5-Pro          &\textbf{67.59}&\textbf{63.10}&\textbf{3.88}&93.80&65.93&3.47&45.52&41.17&3.46&85.62&\textbf{71.40}&3.84 &68.93\\
& GPT-4o                  &57.44&52.08&3.74&99.60&69.59&3.42&44.09&47.87&3.36&83.11&63.30&3.78 &66.92\\
& GPT-5                   &60.52&51.52&\textbf{3.88}&98.33&69.07&3.50&\textbf{58.69}&\textbf{53.23}&\textbf{3.62}&\textbf{88.07}&69.89&\textbf{3.93} &\textbf{70.66}\\
& Llama-3.1-70B-Inst      &42.91&40.32&3.46&84.97&62.58&3.23&23.35&17.76&3.19&58.70&41.74&3.55 &53.41\\
& Llama-3.1-8B-Inst       &44.30&46.73&3.33&90.83&68.77&2.95&18.62&27.65&2.67&63.60&58.22&3.41 &55.49\\
& Qwen2.5-72B-Inst        &55.93&54.60&3.58&99.53&70.26&3.32&52.22&47.91&3.44&83.69&62.50&3.69 &67.27\\
& Qwen2.5-7B-Inst         &41.14&45.69&3.33&96.13&67.89&3.22&29.38&26.79&3.11&61.24&54.09&3.41 &56.98\\
\midrule
\multirow{9}{*}{LLM-Based}
& DeepSeek-V3             &43.79&45.57&3.42&97.87&66.07&3.36&31.14&34.28&3.07&55.13&52.10&3.51 &57.76\\
& Gemini-2.5-Flash        &52.65&50.82&3.42&92.33&66.71&3.28&54.24&44.33&3.23&72.13&61.48&3.58 &63.74\\
& Gemini-2.5-Pro          &53.06&49.61&3.45&82.73&58.76&3.32&50.88&38.86&3.35&74.11&61.44&3.60 &61.99\\
& GPT-4o                  &45.97&44.55&3.52&98.33&69.42&3.32&41.50&38.09&3.03&66.98&54.11&3.57 &60.65\\
& GPT-5                   &48.64&45.71&3.71&96.73&67.73&\textbf{3.53}&49.74&44.91&3.51&77.29&64.04&3.81 &65.50\\
& Llama-3.1-70B-Inst      &31.85&37.47&3.10&73.92&63.18&3.02&12.47&12.09&2.82&36.70&32.34&2.97 &44.85\\
& Llama-3.1-8B-Inst       &24.05&35.61&3.29&22.29&22.99&3.18&7.96&13.53&2.54&29.68&34.58&2.94 &35.81\\
& Qwen2.5-72B-Inst        &43.18&44.58&3.45&96.67&68.56&3.26&36.06&30.39&3.28&57.38&50.41&3.43 &57.97\\
& Qwen2.5-7B-Inst         &30.84&37.05&2.96&60.94&69.86&2.67&12.00&12.60&2.76&37.17&43.27&2.76 &43.89\\
\midrule
\multirow{6}{*}{MLLM-Based}
& Gemini-2.5-Flash        &39.27&40.57&3.39&96.13&69.46&3.30&36.05&31.42&3.33&58.09&49.73&3.51 &57.61\\
& Gemini-2.5-Pro          &40.26&40.82&3.45&86.13&61.32&3.32&36.79&30.54&3.34&63.15&51.47&3.57 &57.01\\
& GPT-4o                  &25.87&31.14&3.38&98.20&\textbf{70.74}&3.27&26.95&31.65&3.12&49.93&42.83&3.49 &53.54\\
& GPT-5                   &32.75&37.35&3.65&91.47&63.05&3.52&34.57&32.98&3.56&61.21&49.32&3.80 &57.77\\
& Qwen2.5-VL-72B-Inst     &29.87&34.23&3.27&93.47&69.46&3.15&25.11&26.57&3.17&42.33&37.24&3.20 &51.17\\
& Qwen2.5-VL-7B-Inst      &17.02&24.28&2.30&58.01&39.83&2.43&5.17&4.24&2.43&10.84&11.26&2.23 &29.87\\
\midrule
\multirow{9}{*}{Rule-Based}
& DeepSeek-V3             &26.05&29.05&3.16&30.70&22.80&3.13&15.53&24.02&2.83&17.83&22.39&2.83 &35.61\\
& Gemini-2.5-Flash        &22.55&27.39&3.10&40.33&29.99&3.03&11.58&22.47&2.84&17.59&20.11&2.97 &35.90\\
& Gemini-2.5-Pro          &31.22&31.57&3.10&44.40&32.45&3.03&14.10&25.18&2.89&27.49&26.51&3.14 &39.68\\
& GPT-4o                  &21.49&26.18&3.17&34.80&26.18&3.17&9.82&18.56&2.81&13.93&17.60&2.97 &34.25\\
& GPT-5                   &31.43&30.73&3.32&21.20&14.98&3.23&15.59&22.81&3.10&23.43&22.60&3.30 &36.81\\
& Llama-3.1-70B-Inst      &15.52&21.39&2.84&36.20&26.84&2.91&14.16&16.06&2.70&13.69&13.28&2.80 &31.85\\
& Llama-3.1-8B-Inst       &21.39&25.45&2.61&34.40&24.68&2.62&9.92&10.68&2.48&14.50&16.51&2.56 &30.24\\
& Qwen2.5-72B-Inst        &25.89&27.86&3.17&36.20&26.74&3.14&10.28&16.92&2.88&16.12&19.21&2.91 &35.10\\
& Qwen2.5-7B-Inst         &25.09&28.61&2.86&38.40&28.33&2.85&11.17&14.15&2.60&11.57&12.40&2.65 &32.41\\
\bottomrule
    \end{tabular}%
    }
\end{table*}

\subsubsection{Implementation and Generation Settings}

All methods use the same retrieval and generation budget. 
We use BGE-M3 as the shared embedding backbone for initial retrieval, so that the comparison focuses on how evidence units are represented before indexing rather than on differences in retrieval encoders. 
The baselines encode evidence in a flat manner, whereas HAM-RAG augments each text or image unit with hierarchy-aware contextual signals from the source document.
For hybrid retrieval, dense and sparse scores are combined with weights 0.7 and 0.3, respectively. 
Remote generators are called through the same OpenAI-compatible interface with temperature 0.2, streaming disabled, and a maximum generation length of 4,000 tokens.

For downstream generation, each method is allowed to use up to 15 retrieved text units and 10 images. 
This budget balances evidence coverage and inference feasibility: some questions require multiple text units to recover sufficient context, while image-intensive questions require several visual references to produce complete interleaved answers. Due to the relatively large scale and higher inference cost of the \textit{Recipe} subset, we randomly sample 500 QA pairs from this subset for evaluation using seed 36, while keeping the same retrieval-and-generation settings as the other subsets.

Invalid generations are counted as failures, retained in the evaluation, and assigned zero scores on the corresponding automatic metrics, so the final results reflect both answer quality and operational robustness.

\subsubsection{Evaluation Metrics}

We evaluate both retrieval quality and multimodal generation quality. 
All methods are evaluated with the same prompts, scripts, and retained failed-case policy.

For retrieval evaluation, we separately assess textual evidence coverage and visual evidence coverage. 
For text retrieval, we report Text Context Recall@15 (TCR@15), an LLM-based set-level metric that measures whether the top-15 retrieved text chunks collectively provide sufficient evidence for answering the query. 
For image retrieval, we report Image Recall (IR), which measures whether the reference images needed by the answer are covered by the retained image set.

For generation evaluation, the main table focuses on the metrics that directly reflect interleaved multimodal answer fidelity: image grounding, local text-image alignment, and overall answer quality. 
We use Image F1 to measure whether the generated answer selects the correct supporting images. 
To further evaluate image placement without introducing an additional threshold or human annotation, we define Image Context BERTScore (Img-CBS), a placement-sensitive continuous metric that applies BERTScore~\citep{zhangbertscore} to compare the text surrounding each correctly selected image in the generated answer with the corresponding local context in the reference answer. 
Unlike Image F1, which only evaluates image selection, Img-CBS rewards a method only when the selected image is placed near semantically appropriate supporting text.
Finally, we use an LLM-based Overall Quality score to assess the correctness, completeness, grounding, and text-image coherence of the generated multimodal answer as a whole.
In Table~\ref{tab:overall_performance}, the MM Avg. column reports the unweighted mean over the four datasets and three multimodal generation metrics: Image F1, Img-CBS, and Qual. \(\times 20\). 

\begin{table}[t]
    \centering
    \small
    \setlength{\tabcolsep}{4.5pt}
    \caption{HAM-Bench retrieval results.}
    \label{tab:retrieval_results}
    \resizebox{\columnwidth}{!}{%
    \begin{tabular}{llcccc}
    \toprule
    Stage & Method / Pipeline & Wukong $\uparrow$ & Wiki $\uparrow$ & arXiv $\uparrow$ & Recipe $\uparrow$ \\
    \midrule
    \multirow{2}{*}{Text (TCR@15)}
    & HAM-RAG & \textbf{0.99} & \textbf{0.80} & \textbf{0.77} & \textbf{0.97} \\
    & Shared Baseline (LLM / MLLM / Rule) & 0.80 & 0.70 & 0.62 & 0.91 \\
    \midrule
    \multirow{3}{*}{Image (IR)}
    & HAM-RAG & \textbf{0.83} & \textbf{1.00} & 0.86 & \textbf{0.97} \\
    & MLLM-Based (CLIP-filtered) & 0.50 & 0.98 & 0.57 & 0.73 \\
    & Shared Baseline (LLM / Rule) & 0.65 & 0.98 & \textbf{0.99} & 0.92 \\
    \bottomrule
    \end{tabular}}
\end{table}

\subsection{Overall Performance}
\label{ssec:overall_performance}

Table~\ref{tab:retrieval_results} summarizes retrieval performance. Since all baselines share the same text retriever, their TCR@15 results are reported jointly; MLLM-Based is separated for image retrieval because it applies CLIP filtering. HAM-RAG achieves higher TCR@15 on all four subsets and higher IR on three. On arXiv, the shared baseline reaches 0.99 IR versus 0.86 for HAM-RAG, likely because flat retrieval retains more figures in this image-dense subset. Since IR measures coverage rather than placement, this exception does not contradict HAM-RAG's stronger local text-image alignment.

Table~\ref{tab:overall_performance} demonstrates that HAM-RAG's advantage is consistent across generator families rather than being driven by a single high-capacity model. For all nine matched backbones, HAM-RAG achieves a higher MM Avg. than its LLM-Based counterpart. Averaged across these backbones, MM Avg. increases from 54.68 to 64.12, corresponding to a 17.3\% relative improvement. This consistency across proprietary and open-source models of different scales indicates that the gain primarily comes from hierarchy-aware evidence organization before generation, rather than model-specific generation capability.

The improvements are most informative on metrics that evaluate whether visual evidence is selected and used in the correct local context. On Wukong, where answers depend on ordered walkthrough steps and interleaved screenshots, HAM-RAG increases the best Img-CBS from 50.82 to 63.10, a 24.2\% improvement over the strongest non-hierarchical baseline. The gains on Recipe and arXiv further indicate that structural context helps associate images with the relevant procedure or section instead of merely retrieving topically related figures. Wiki is less discriminative for image selection because each question contains only one reference image and Img-F1 frequently saturates; its results should be interpreted together with Img-CBS and Qual. Overall, the results show that hierarchy improves not only aggregate performance, but also visual grounding and local text-image alignment.

In summary, the results support our central claim that document hierarchy is a critical signal for multimodal RAG over structured documents. Compared with non-hierarchical paradigms, HAM-RAG better preserves source-document context and improves the alignment between retrieved images and their corresponding textual units, which is essential for faithful interleaved text-image generation.

\subsection{Ablation Study}

\begin{figure}[t]
    \centering
    \includegraphics[width=\columnwidth]{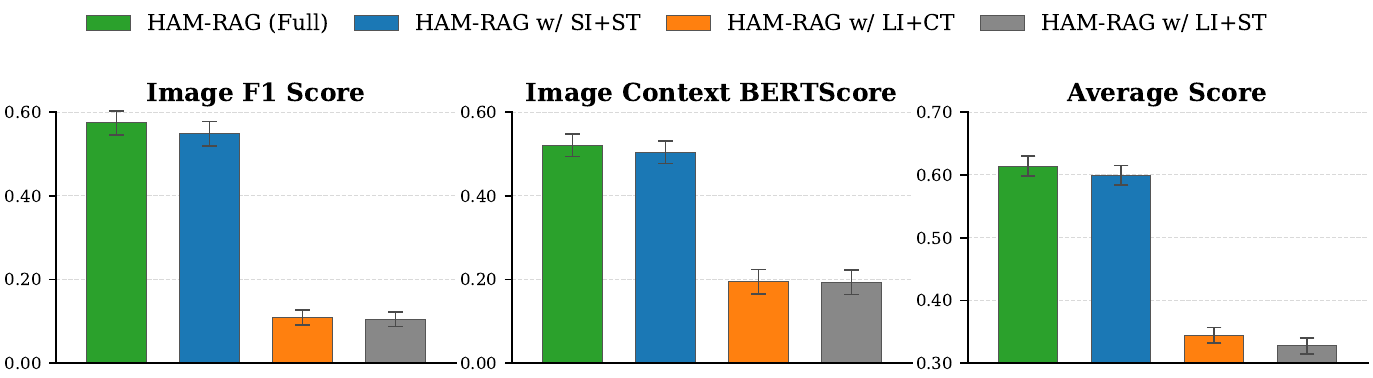}
    \captionsetup{justification=justified,singlelinecheck=false}
    \caption{HAM-RAG ablation results on Wukong for Image F1 Score, Image Context BERTScore, and Average Score.}
    \Description{Three bar charts comparing the full HAM-RAG model with SI+ST, LI+CT, and LI+ST variants on 540 Wukong questions. The metrics are Image F1 Score, Image Context BERTScore, and a three-metric Average Score. Bar heights represent means, and error bars represent 95 percent confidence intervals.}
    \label{fig:ablation}
\end{figure}

Using GPT-4o on 540 Wukong questions, we separately ablate the hierarchy encoded in image and text representations. SI and LI denote the context-aware and weaker local image representations, respectively; LI omits context-aware visual descriptions. CT and ST denote hierarchy-aware and isolated text representations. SI+ST weakens text hierarchy, LI+CT weakens image hierarchy, and LI+ST weakens both. Figure~\ref{fig:ablation} reports means with 95\% confidence intervals (\(\bar{x}\pm1.96s/\sqrt{540}\)); Average Score is the per-question mean of Image F1, Img-CBS, and Qual./5. The full model has the highest mean on all three metrics. Replacing SI with LI sharply reduces both image metrics. SI+ST yields smaller declines, whereas LI+ST has the lowest Average Score, showing complementary image- and text-side contributions.

\subsection{Robustness and Cost}
\label{ssec:robustness_cost}

{\tolerance=3000\emergencystretch=1em\hyphenpenalty=10000\exhyphenpenalty=50
Across 57,123 main-experiment generation runs, the four evaluated frameworks produced only 38 failures, corresponding to a failure rate of approximately 0.067\%. 
Under the same evaluation setting, HAM-RAG produced only 2 failures, fewer than LLM-Based (29) and MLLM-Based (7), with both failures occurring on the \textit{arXiv} subset where long section context and dense figures make prompt construction more difficult. 
The Rule-Based pipeline produced no failures, but it uses a much lighter text-first generation and post-hoc image insertion process rather than generating fully grounded interleaved multimodal answers. 
Macro-averaged across the four subsets under the GPT-4o pricing model, HAM-RAG has an estimated cost of 3.578 cents per question, lower than LLM-Based and MLLM-Based but higher than Rule-Based. 
This pattern reflects the intended trade-off of structure-aware multimodal RAG: preserving document organization and local text-image evidence requires richer prompts, but yields stronger grounding and a very low failure rate.
\par}

\section{CONCLUSION}

{\tolerance=3000\emergencystretch=1em\hyphenpenalty=10000\exhyphenpenalty=50
This paper studies faithful interleaved generation over structured multimodal documents, where flat retrieval weakens dependencies among sections, local contexts, and visual evidence.
HAM-RAG addresses this issue by incorporating multi-level document context into text and image representations, preserving source organization and local text-image relations across retrieval and generation.
HAM-Bench further provides a unified benchmark across diverse structured-document scenarios.
Experiments across datasets and backbones show consistent gains in image grounding, local text-image alignment, and holistic answer quality, confirming document hierarchy as a key grounding signal for coherent multimodal generation.
These findings support reliable assistants for industrial SOPs, maintenance guides, and operational manuals, where answers must remain faithful to procedural structure and local text-image evidence.
\par}

\bibliographystyle{apiems}
\renewcommand{\refname}{REFERENCES}
\bibliography{reference}

\end{document}